\documentclass[twocolumn,preprintnumbers,amsmath,amssymb,superscriptaddress]{revtex4-2}

\usepackage{graphicx}
\usepackage{dcolumn}
\usepackage{bm}

\begin{document}

\title{Chemical logic gates on active colloids}
\author{Jiang-Xing Chen}
\altaffiliation{jxchen@hdu.edu.cn}
\affiliation{Department of Physics, Hangzhou Normal University, Hangzhou 311121, China}
\author{Jia-Qi Hu}
\affiliation{Department of Physics, Hangzhou Normal University, Hangzhou 311121, China}
\author{Raymond Kapral}
\altaffiliation{r.kapral@utoronto.ca}
\affiliation{Chemical Physics Theory Group, Department of Chemistry, University of Toronto, Toronto, Ontario M5S 3H6, Canada}

\begin{abstract}
Synthetic active colloidal systems are being studied extensively because of the diverse and often unusual  phenomena these nonequilibrium systems manifest, and their potential applications in fields ranging from biology to material science. Recent studies have shown that active colloidal motors that use enzymatic reactions for propulsion hold special promise for applications that require motors to carry out active sensing tasks in complicated biomedical environments. In such applications it would be desirable to have active colloids with some capability of computation so that they could act autonomously to sense their surroundings and alter their own dynamics to perform specific tasks. Here we describe how small chemical networks that make use of enzymatic chemical reactions on the colloid surface can be used to construct motor-based chemical logic gates. Some basic features of coupled enzymatic reactions that are responsible for propulsion and underlie the construction and function of chemical gates are described using continuum theory and molecular simulation. Examples are given that show how colloids with specific chemical logic gates can perform simple sensing tasks. Due to the diverse functions of different enzyme gates, operating alone or in circuits, the work presented here supports the suggestion that synthetic motors using such gates could be designed to operate in an autonomous way in order to complete complicated tasks.
\end{abstract}

\maketitle

\section{Introduction}
Extensive investigations of colloidal motors with micrometer dimensions that use various mechanisms to produce propulsion have been carried out, and review articles summarize much of this research.~\cite{wangbook2013,kapral2013,sanchez14,Pumera2015,zoettl2016,Illien2017,roadmap2020,DWDYMS15} Growing interest in such active agents stems from their broad potential uses as vehicles for drug delivery, cargo transport, motion-based species detection, active self-assembly, micro fluidics, medical as well as other applications.~\cite{DWDYMS15,MHS15,wang2019application,xu2019self,tejeda2019virus,soto2019onion,ying2019micro} In order for such micromotors to perform useful tasks effectively, their directed motion must be controlled in some way, even in the presence of strong thermal fluctuations. Chemical gradients, walls and external fields, among other means, have been used to guide their motions.~\cite{PNASW2013,SenRev2013,Suzanne14,Li2015,Das15,Simmchen16,IL16,PUDD18,Jizhuang2018,CRRK18,Guo18} Rather than seeking to externally direct motor motion, it would be desirable if the motors themselves could discover ways of responding to stimuli to achieve specific goals.

Chemically-powered colloidal motors can be propelled using diffusiophoretic mechanisms~\cite{anderson1983,golestanian2007,PUD16,OPD17,GK18a}, and in this context motors that make use of enzymatic chemical reactions on the colloid surface~\cite{Dey2015,MHPS16,MHS16,ZGMS18,ARFPOS19,MJAHMSS15} are the focus of this work. Such enzyme-powered micro and nanomotors have important properties, such as biocompatibility, versatility, and fuel bioavailability, that make them attractive for applications.~\cite{MJAHMSS15,ARFPOS19,Toebes19,PPJLSSRS19,Sanchez-gate2019,Tang20} Usually the chemical fuel that powers motor motion is supplied directly by the environment; however, active colloids that make use of coupled enzymatic reactions for propulsion and multi-fueled enzymatic motors have been studied in the laboratory.~\cite{Wang2013,Stadler2015,Wilson2016,Stadler2017} Some fundamental aspects of the operation of such motors remain unexplored, including the structure of the colloidal catalytic surface and characteristics of substrate species involved in the coupled reactions, the conversion rate of fuel in surface enzyme networks, and effects related to fuel supply.

A body of earlier research has shown how DNA, RNA and protein networks can be used to construct chemical logic gates, and how circuits built from these gates could be used to carry our simple computations.~\cite{AR94,UM2006,PSSPK2008,Prasanna2006,MRDI2013,P13,KPS2017,CKH20,Seelig2013}  For example, a logic network composed of three enzymes operating in concert as four concatenated logic gates (AND/OR) was designed to process four different chemical input signals and finally produces a pH change as the output signal.~\cite{Privman09} Also, various two-input gates built from de novo-designed proteins have been proposed recently~\cite{Chen19}, contributing to the design of programmable protein circuits.~\cite{CHEN2021}

Given the substantial amount of research on the construction of protein and other chemical logic gates and circuits, it should be possible to exploit this research to construct programmable enzyme-powered motors. Colloids with linear dimensions of one to a few micrometers can support tens to hundreds of thousands of enzymes on their surfaces; thus, one can construct colloids coated with several different enzymes to implement chemical logic gate functions as described above. In this connection, micromotors with a gated pH responsive DNA nanoswitch have been made and studied in the laboratory to function as on-demand payload delivery systems.~\cite{PPJLSSRS19,Sanchez-gate2019}

In this paper we investigate various ways in which small chemical networks that make use of enzymatic chemical reactions on the colloid surface can be used to construct motor-based chemical logic gates. In this way the motors may perform chemical computational tasks that allow them to control their dynamics by sensing the characteristics of the environment in which they move. Below we present a discussion of some basic aspects of how coupled enzyme reactions influence colloid propulsion, and provide examples of how specific chemical logic gates can be implemented to allow a colloidal motor to sense and respond to its environment in different ways, thus, performing simple tasks.

\section{Chemical gates and motors with coupled enzymatic reactions}\label{sec:coupled-reactions}
Coupled enzymatic reactions involving glucose oxidase and catalase enzymes coated on a Janus surface have been studied experimentally~\cite{Stadler2015,Wilson2016,Stadler2017}. In this system the reaction of glucose (Glc) in the presence of $O_2$ catalysed by the enzyme glucose oxidase $(GO_x)$ yields gluconic acid (GlcA) and $H_2O_2$, ${\rm Glc} +O_2 \stackrel{GO_x}{\rightarrow} H_2O_2 +{\rm GlcA}$. This reaction, in turn, supplies the $H_2O_2$ fuel that is used by catalase in the reaction $H_2O_2 \stackrel{Cat}{\rightleftharpoons} H_2O +O_2$ to propel the colloid. Since $H_2O_2$ is not a desirable environmental species in biochemical applications it is advantageous to supply fuel directly on the catalytic surface where it is used, instead of the more conventional global fuel supply; thus, the chemical kinetics of systems with locally supplied fuel merit detailed study.

\begin{figure}[htbp]
\centering
\resizebox{0.8\columnwidth}{!}{
      \includegraphics{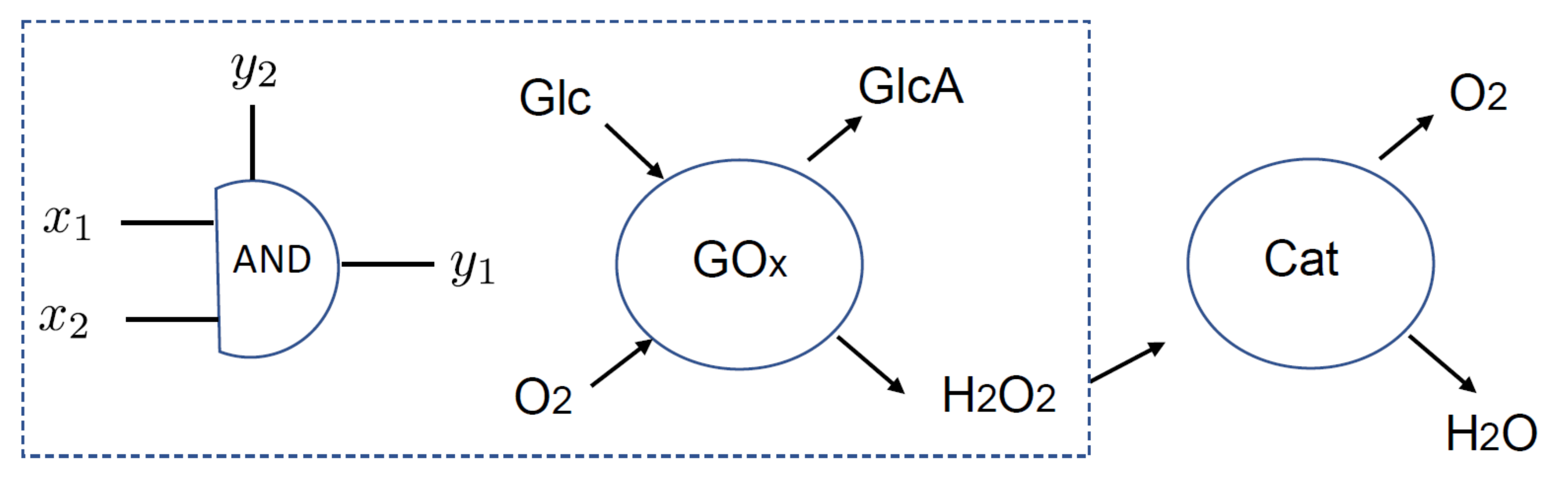}}
        \caption{
       A small protein network that simulates an AND gate whose output is a species that is fuel for a second enzyme. The glucose and oxygen substrates correspond to the inputs $x_1$ and $x_2$ to the AND gate while hydrogen peroxide is the desired output $y_1$. The product $y_2={\rm GlcA}$ is not monitored. The hydrogen peroxide is then processed by the catalase enzyme to produce $O_2$ and $H_2O$.
    } \label{fig:AND-gate}
\end{figure}

The properties of colloidal motors that support chemical logic gates involve such coupled enzymatic reactions on the colloid surface, and some of these reactions may be used to implement to sensing tasks, while others are responsible for motor motion. Indeed, Fig.~\ref{fig:AND-gate} shows how the reaction catalysed by $GO_x$ can be mapped onto an AND gate. The inputs to the gate are $x_1={\rm Glc}$ and $x_2=O_2$, while the target output is hydrogen peroxide, $y_1=H_2O_2$. The GlcA product is not monitored. The reaction models an AND gate since both substrates are required for product formation.~\cite{binary-gates}

A simple version of such coupled chemical reactions on the colloid surface can serve to illustrate the roles of various factors, such as reaction rates, manner of fuel supply, and enzyme distributions on the colloid dynamics. For a spherical colloid with radius $R$ we suppose that on the upper hemisphere ($H_u$) a fraction $f^u_1$ of surface area is covered by enzyme $E_1$, randomly distributed on that hemisphere, while the remaining fraction $f^u_2=1-f^u_1$ is covered by enzyme $E_2$ (see sketch in fig. 2(a)). On the lower hemispherical cap ($H_\ell$) a fraction $f^\ell_1$ is covered by enzyme $E_1$, while on the remainder of the hemispherical surface the fraction $f^\ell_0=1-f^\ell_1$ sites are inactive enzymes $E_0$ or empty sites.

Enzyme 1 catalyzes the reaction
\begin{equation}
A +R_1^*\underset{E_1}{\stackrel{\kappa_1}{\rightarrow}} I +P_1^*, \quad {\rm on \; enzyme \; 1 \; surface \; fractions}, \label{eq:model1a}
\end{equation}
where species with a superscript $\ast$ are assumed to be held constant by reservoirs or are in excess. The fixed concentration $c^*_{R_1}$ of $R_1^*$ is incorporated in the rate constant $k_1$ and $\kappa_1=k_1/(4\pi R^2)$ is the rate constant per unit surface area. (If $c^*_{R_1}$ is in excess or above a threshold the reaction~(\ref{eq:model1a}) is controlled by the concentration of $A$. In this case, one input is always on, and the reaction can act effectively as a buffer gate that produces species $I$ if $A$ is an input.) Enzyme 2 catalyzes the reaction,
\begin{equation}
I \underset{E_2}{\stackrel{\kappa_2}{\rightarrow}} B, \quad {\rm on \; enzyme \; 2 \; surface \; fractions}\label{eq:model1b}
\end{equation}
with $\kappa_2=k_2/(4\pi R^2)$. In these reactions we see that the product $I$ produced on the motor surface by reaction (\ref{eq:model1a}) serves as the fuel for the motor reaction (\ref{eq:model1b}). In addition, we assume that the species $B$ and $I$ are degraded in the bulk phase and ultimately produce $A$. (Alternatively, one can suppose that $B$ and $I$ are removed from the system by catabolic reactions and $A$ is supplied by reservoirs.) Thus, the reactions $B \stackrel{k^B_b}{\rightarrow} A$ and  $I \stackrel{k^I_b}{\rightarrow} A $, take place in the fluid phase with bulk phase rate constants $k^B_b$ and $k^I_b$, and maintain the system in a nonequilibrium state.

This scheme could also model reactions catalysed by other enzymes; for example, the enzyme choline oxidase (CHO) that catalyzes the reaction of choline (Ch) and $O_2$ to give betaine  aldehyde (Be) and $H_2O_2$, ${\rm Ch} +O_2 \stackrel{ChO}{\rightarrow} H_2O_2 +{\rm Be}$.~\cite{KPS2017}

Instead of a single catalytic reaction, the reaction network using both Eqs.~(\ref{eq:model1a}) and (\ref{eq:model1b}) also can model an AND gate using the coupled reactions of two enzymes such as $GO_x$ and horseradish peroxidase (HRP). The $H_2O_2$ from the $GO_x$ catalysis is then processed by HRP in the reaction, $H_2O_2 +{\rm ABTS}  \stackrel{HRP}{\rightarrow} O_2 +{\rm ABTS}_{ox}$. With the $O_2$  $(R_1^*$ above) concentration fixed, one can take as inputs $x_1={\rm Glc}$ and $x_2={\rm ABTS}$, where ABTS is 2,2'-azino-bis(3-ethylbenzothiazoline-6-sulfonic acid) and the output of the AND gate is ${\rm ABTS}_{ox}$, the oxidized form of ABTS.~\cite{KPS2017}

We use continuum theory and molecular simulation to investigate the properties of coupled enzyme kinetics on colloids based on Eqs.~(\ref{eq:model1a}) and (\ref{eq:model1b}).

\begin{figure}[htbp]
\centering
\resizebox{1.06\columnwidth}{!}{
      \includegraphics{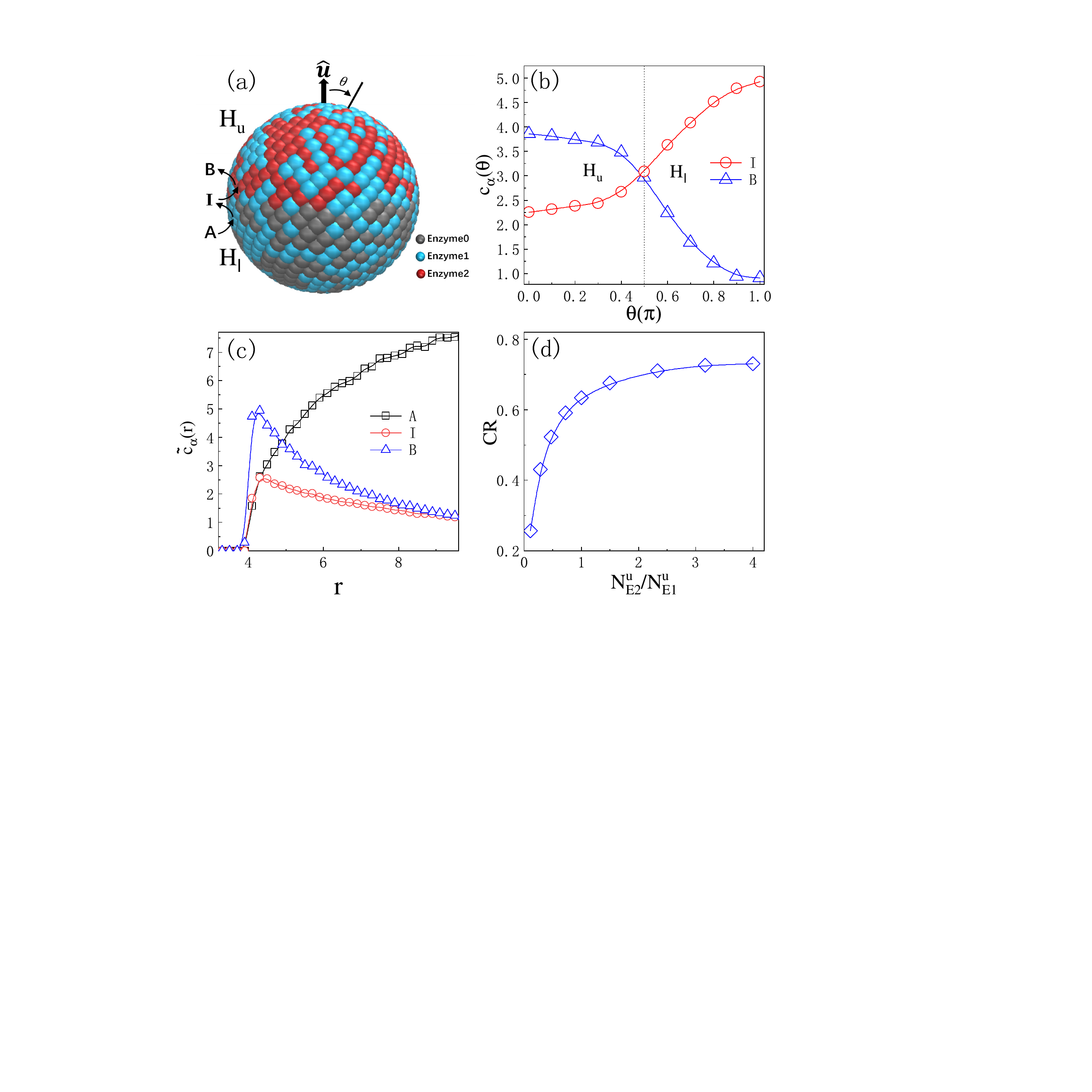}}
        \caption{
       (a) A colloidal Janus motor with radius $R=4.0$ and 1000 coarse-grained enzymatic sites on its surface. The fractions of $E_1$ sites (blue) on the two hemispheres are $f^u_1=f^\ell_1=1/2$, so that $E_1$ and $E_2$ each occupy 250 sites on the $H_u$ cap, while there are 250 $E_1$ and 250 $E_0$ sites (grey) on the $H_\ell$ cap. The motor axis is defined by the unit vector $\hat{\bm{u}}$ pointing from the center-of-mass of the colloid to that of the $H_u$ cap containing $E_2$ sites (red). The energy parameters are $\epsilon_A=\epsilon_I=1.0$ and $\epsilon_B=0.1$.  (b) The intermediate fuel species $I$ and product $B$ concentrations $c_\alpha(\theta)$ versus $\theta$ at $r$=5.0. (c) The $H_u$-cap radial concentrations $\tilde{c}_{\alpha}(r)$ for three substrate species, $A$ (triangles), $I$ (squares), and $B$ (circles). (d) The dependence of conversion ratio $CR$ on $N^u_{E_2}/N^u_{E_1}$, for $f^\ell_1=0$ so that the $H_\ell$ cap is inactive and $N^u_{E_2}+N^u_{E_1}=500$.} \label{fig:local-fuel}
\end{figure}

\vspace{0.25cm} \noindent
{\sf Continuum model}: The velocity of the motor can be computed using the formula for the diffusiophoretic propulsion velocity~\cite{anderson1983,golestanian2007,PUD16,OPD17,GK18a} to give~\cite{stick}
\begin{eqnarray}\label{eq:colloid-prop-velocity}
\bm{V}_u = V_u \hat{\bm{u}} =\frac{k_B T}{\eta} \Big[ \Lambda_{IA}\overline{\bm{\nabla}_s c_I(\bm{r})}^S + \Lambda_{BA}\overline{\bm{\nabla}_s c_B(\bm{r})}^S  \Big]
\end{eqnarray}
where the overline denotes an average over the surface of the colloid, and $\Lambda_{\alpha \alpha'}=\int_R^{R+r_c} dr \; (r-R) \big(e^{-\beta V_{\alpha c}} -e^{-\beta V_{\alpha' c}} \big)$
accounts for the interactions of solute molecules of type $\alpha$ with the colloid through repulsive potentials $V_{\alpha}$ with finite range $r_c$. The velocity is directed along the unit vector $\hat{\bm{u}}$ that points from the $H_\ell$ to $H_u$ hemispheres as shown in Fig.~\ref{fig:local-fuel} (a). The numerical values of the motor velocity can be found by substituting the solutions of the reaction-diffusion equations, subject to boundary conditions on the motor surface, into Eq.~(\ref{eq:colloid-prop-velocity}). This calculation is given in Appendix~\ref{app:cont-model1}. Among other factors, the propulsion velocity depends on the rate constants $k_1$ and $k_2$, and the fractional coverage.

By contrast, if the fuel $I$ is supplied directly from the fluid phase, as is usually done for diffusiophoretic motors, we have
\begin{equation}\label{eq:prop-velocity-bulk-fuel}
\bm{V}_u =\frac{k_B T}{\eta}  \Lambda_{IB}\overline{\bm{\nabla}_s c_I(\bm{r})}^S .
\end{equation}
The full expression is also given in Appendix~\ref{app:cont-model1}.

Comparisons of Eqs.~(\ref{eq:colloid-prop-velocity}) and (\ref{eq:prop-velocity-bulk-fuel}) allow one to quantify the differences between fuel supply at the motor surface and supply from the boundaries. Not only do two rate constants, $k_1$ and $k_2$, appear when fuel is supplied locally, but the bulk phase reactions and boundary conditions couple all three concentrations, $c_A$, $c_B$ and $c_I$. Hence, surface fuel production leads to a rich structure that could be exploited control motor motion.

\vspace{0.25cm} \noindent
{\sf Simulation}: Simulation of the motor dynamics is based on a coarse-grained microscopic description of the entire system. The surface of the colloid is covered by spherical beads that act as coarse-grained groups of enzymes. The $E_1$  and $E_2$ enzyme groups are uniformly and randomly distributed on the colloid surface  with fractions as described above and sketched in Fig.~\ref{fig:local-fuel} (a). The reactions on the surface of the colloid take place on the chemically active enzymatic sites. The strengths of the repulsive colloid-solute potentials are gauged by $\epsilon_\alpha$  energy parameters, and for these simulations $\epsilon_A=\epsilon_I > \epsilon_B$. The evolution of the system is carried out using hybrid molecular dynamics-multiparticle collision dynamics~\cite{MK99,MK00,Ray08,GIKW09} Further information on the simulation model and parameters are given in Appendix~\ref{app:sim-model1}.

\vspace{0.25cm} \noindent
{\sf Concentration fields and motor velocity}:
From Eq.~(\ref{eq:colloid-prop-velocity}) one can see that the concentration gradient fields in the colloid vicinity play an important part in the diffusiophoretic mechanism, and, for the sequential reaction mechanism  of interest here, the results of simulations and continuum calculations allow one to assess the factors giving rise to the spatial structure of these fields. Specifically, the gradient fields $\partial_\theta c_B(R, \theta)$ and $\partial_\theta c_I(R, \theta)$ enter the expression for the $V_u$ in Eqs~(\ref{eq:colloid-prop-velocity}) and (\ref{eq:Vd-simple-model-2}). Plots of $c_B(5,\theta)$ and $c_I(5,\theta)$ versus $\theta$ are shown in Fig.~\ref{fig:local-fuel} (b) for a radius value $r=5$, somewhat outside the boundary region at $r_c=4.125$ where the potentials act. The $c_B$ and $c_I$ profiles are typical of Janus colloids where the largest gradient is in the vicinity of the equator where the $H_u$ and $H_{\ell}$ caps meet. One also observes the depletion of locally-supplied I on the upper hemisphere. The pronounced asymmetry of these fields around the colloid indicates that local fuel can provide effective self-propulsion.

The overall structure of the concentration fields of all three species, $\alpha = A,I,B$ near the $H_u$ cap is seen in the plots of the radial concentrations, $\tilde{c}_{\alpha}(r)$, in Fig.~\ref{fig:local-fuel} (c). These radial concentrations were constructed from averages of the species density fields only over the angles corresponding to the $H_u$ cap. The plots show that $I$ is efficiently converted to product $B$ in the motor vicinity, while species $A$ remains in high concentration farther from the colloid. Thus, the colloid is able to function as a motor without high concentrations of the undesirable fuel $I$ in the bulk phase, an important feature for biological applications where undesirable fuels such as $H_2O_2$ are required. Furthermore, the consumption characteristics of the intermediates can have an influence on the design of motor chemical gates based on coupled enzyme networks.

To examine more directly how the locally produced fuel $I$ is converted to product $B$ molecules, we take $f^\ell_1=0$ since then $I$ is produced only on the $H_{u}$ cap that also has $E_{2}$ enzymes. The ratio of the average number of product $B$ molecules, $\bar{N}_B$, to the average number fuel $I$ molecules, $\bar{N}_I$, resulting from reactions on the $H_u$ cap defines the conversion ratio, $CR=\bar{N}_B/\bar{N}_I$. This ratio depends on the fractional occupancies of the $E_1$ and $E_2$ enzymes on the $H_u$ cap,
$N^u_{E_2}/N^u_{E_1}=f^u_2/f^u_1$, since the probability of species $I$ to undergo a surface reaction to $B$ before escaping to the bulk depends on this ratio. This dependence is shown in Fig.~\ref{fig:local-fuel} (d). For $N^u_{E_2}/N^u_{E_1} < 1$, $CR$ increases rapidly and then reaches a plateau value $CR \approx 0.75$. Since the amount of supplied fuel decreases as $N^u_{E_2}/N^u_{E_1}$ increases, this suggests that a fractional coverage of $f^u_1 \approx 1/2$ may be optimal.

\begin{figure}[htbp]
\centering
\resizebox{1.06\columnwidth}{!}{
      \includegraphics{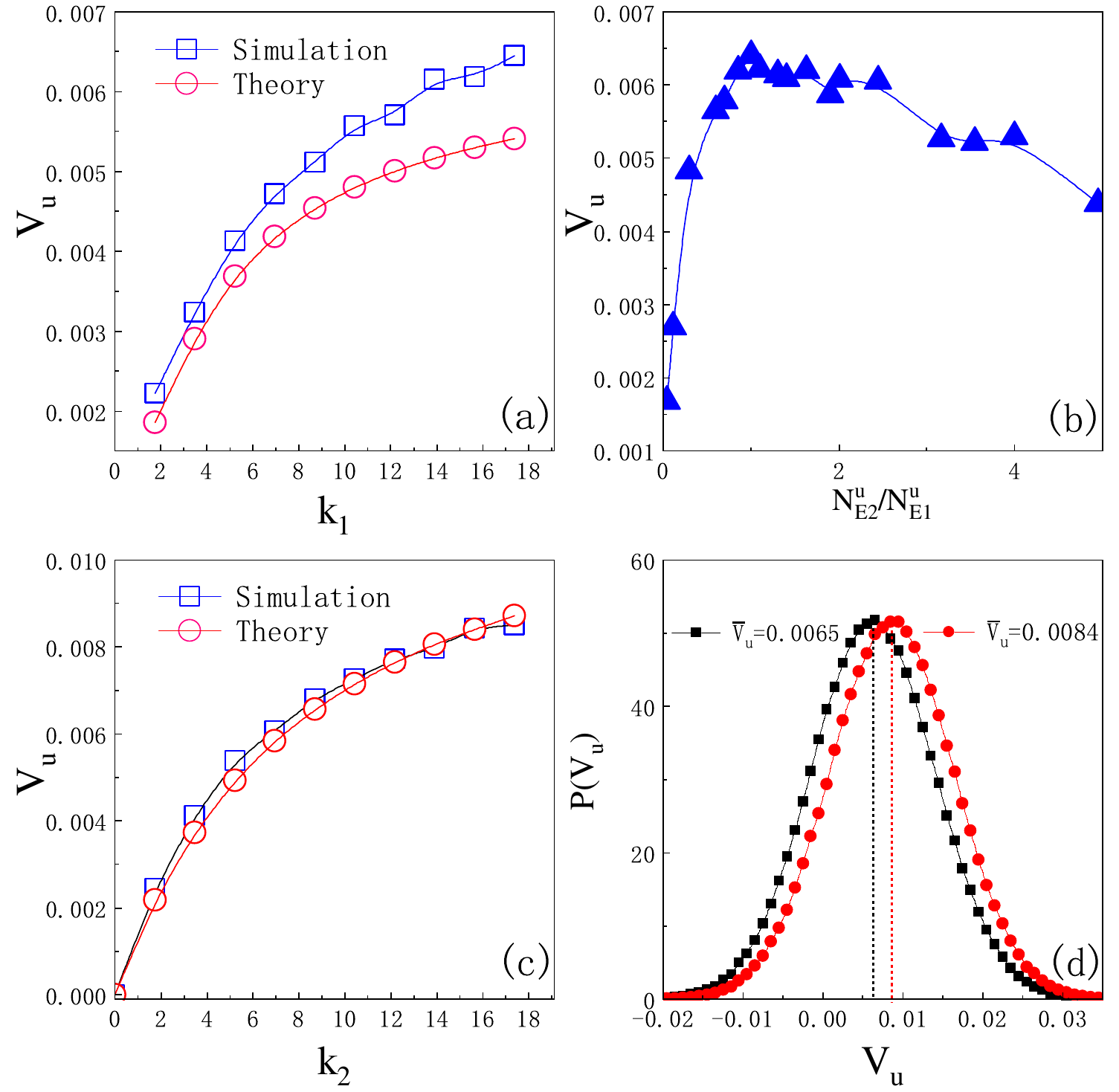}}
        \caption{
       (a) Velocity $V_u$ as a function of the reaction rate constant $k_1$ for fixed $k_2$ for the colloid described in Fig.~\ref{fig:local-fuel} (a). The squares and circles are calculated from simulation and theory, respectively. (b) The dependence of $V_u$ on the ratio $N^u_{E_2}/N^u_{E_1}$. (c) The velocity $V_u$ from simulation (squares) and theory (circles) as a function of $k_2$ when fuel $I$ is supplied from the bulk. (d) Velocity probability distribution $P(V_u)$ versus $V_u$ for local (squares) and global (circles) fuel supply. The data is averaged from ten independent realizations.
    } \label{fig:velocity}
\end{figure}
Results for the propulsion velocity of the motor are given in Fig.~\ref{fig:velocity}. The component of the average motor velocity along $\hat{\bm{u}}$ may be computed in the simulation from $V_u=\left\langle\bm{V}(t)\cdot\bm{\hat{u}}\right\rangle$, where $\bm{V}(t)$ is the instantaneous velocity of the center-of-mass of the colloid and the angular brackets denote an average over time and different realizations of the dynamics. The simulation values are compared to the continuum results using Eqs.~(\ref{eq:colloid-prop-velocity}) and (\ref{eq:Vd-simple-model-2}) in Fig.~\ref{fig:velocity} (a) for different values of $k_1$ and fixed $k_2$. In the simulation $k_1$ is varied by changing the reaction probability $p^R_1$ on $E_1$ with $p^R_2=1.0$ for $E_2$. In the continuum model these rate coefficients are estimated from the kinetic theory expression in Appendix~\ref{app:sim-model1}. (The abscissa values in the plots are the kinetic theory rate coefficients.) The increase in $V_u$ with $k_1$ is expected since more fuel is supplied locally to the $E_2$ enzymatic reaction that powers propulsion; however, the deviation from linear increase can be attributed to the inability of $E_2$ to process the fuel quickly enough before $I$ escapes to the bulk phase. Both simulation and continuum theory are in qualitative agreement and show this effect.

The velocity $V_u$ also depends on fractional occupancy of $E_1$ and $E_2$ on the $H_u$ cap, $f^u_2/f^u_1=N^u_{E2}/N^u_{E1}$ as shown in Fig.~\ref{fig:velocity} (b). Here, in contrast to Fig.~\ref{fig:local-fuel} (d), the $H_\ell$ cap occupancy is still $f^\ell_1=f^\ell_0=1/2$. The rapid increase of $V_u$ with $N^u_{E2}/N^u_{E1}$ reflects the corresponding increase in local fuel supply (seen Fig.~\ref{fig:local-fuel} (d)) on increased numbers of $E_2$ enzymes, while the decrease for larger values of $N^u_{E2}/N^u_{E1}$ is due to the saturation of CR with decreased numbers of $E_1$ enzymes. These results indicate that choice of the optimal fractional occupancy of $E_1$ and $E_2$ gives rise to the most powerful propulsion, another feature that could be exploited when designing motors with coupled enzyme kinetics.

The above results are compared to those where fuel $I$ is directly supplied from the bulk in Fig.~\ref{fig:velocity} (c). The colloid now has a fraction $f^u_2=1/2$ of $E_2$ sites on the $H_u$ cap, while all other sites on the colloid are chemically inactive $E_0$ sites. Only the reaction rate $k_2$ enters this calculation (see reaction (A2) and Eq. (A30) in Appendix A) . The plots are similar to those in panel (a), although the velocity is somewhat larger. This is evident in Fig.~\ref{fig:velocity} (d) where the velocity probability distribution functions $p(V_u$) are plotted. The peaks lie at $V_u=0.0065$ and $V_u=0.0084$ for local and global fuel supply, respectively, and one also sees that thermal fluctuations are strong. Lastly, for global fuel supply simulation and continuum theory are in quantitative agreement, as shown in fig. 3(c). For local fuel supply, the continuum model is not able to capture the full reactive dynamics in the boundary layer since the reaction rates it uses are averages over the entire hemisphere.

\section{Motors with chemical gates that sense their environments}\label{sec:intellegent}

We now show how other chemical logic gates can be used by motors to sense certain aspects of their environments and respond to these chemical signals by carrying out specific tasks. The examples are simple and intended to be illustrative, although the tasks the colloids perform are not very challenging.

\vspace{0.25cm} \noindent
{\sf INH gate:} The first task is to prevent a colloid from being captured by a source. We suppose that the system contains a colloid (C) and a fixed spherical source (S) of a chemical species $I$. The colloid responds to the gradient field of this species through a diffusiophoretic mechanism and may be attracted to the source. We want the motor to sense the magnitude of this concentration field and respond to it in a way that prevents capture by the source. We do this by implementing an inhibitor (IHN) gate on the colloid.

\begin{figure}[htbp]
\centering
\resizebox{0.8\columnwidth}{!}{
      \includegraphics{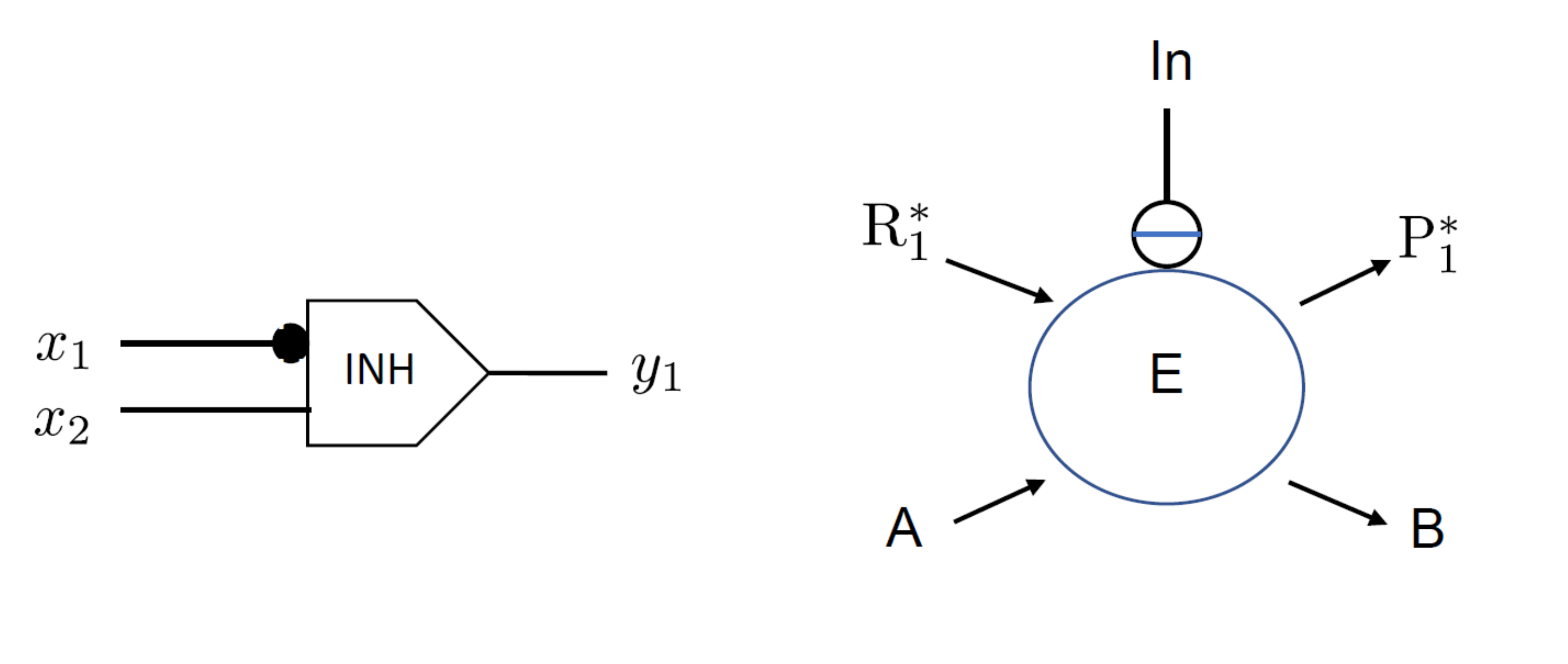}}
        \caption{
       An enzymatic reaction that simulates an inhibitor INH gate. The enzyme E catalyzes the reaction of substrates A and ${\rm R}^*_1$ to form products B and ${\rm P}^*_1$. The open circle with a bar signifies that species In is an inhibitor for the reaction. The substrate A and inhibitor In correspond to the inputs $x_1={\rm In}$ and $x_2={\rm A}$ to the INH gate, while the output is $y_1={\rm B}$.
    } \label{fig:INH-gate}
\end{figure}

Protein logic gates can be constructed using the properties of inhibitors that prevent certain reactions from taking place. For this purpose we consider the reaction scheme in Fig.~\ref{fig:INH-gate} that represents an IHN gate. This gate is modeled on earlier work for a NOT gate that employs inhibitor molecules.~\cite{AR94} In this scheme an enzyme (E) catalyzes the reaction ${\rm R}^*_1 + {\rm A}  \stackrel{E}{\rightarrow} {\rm P}^*_1 + {\rm B}$, where the * on species ${\rm R}^*_1$ and ${\rm P}^*_1$ again signifies that they are held fixed by reservoirs. The presence of an inhibitor (In) prevents the reaction from taking place. (The inhibitor gate can also be constructed from NOT and AND gates.) A colloid with an IHN gate can sense the concentration of inhibitor in the environment and decide to open or close the activity of the enzyme network on the motor. The colloid responds to the state of the gate by changing  its motion.

\begin{figure}[htbp]
\centering
\resizebox{1.05\columnwidth}{!}{
      \includegraphics{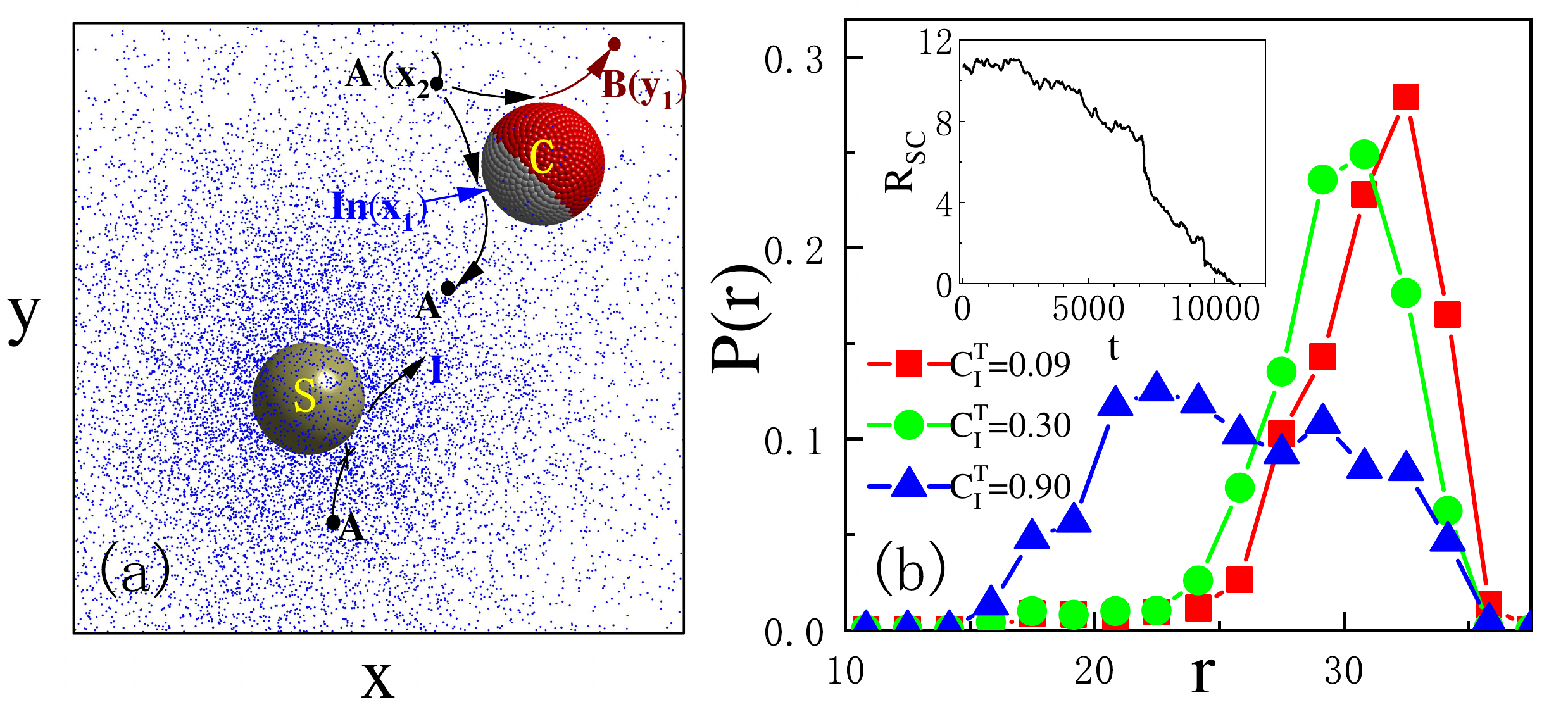}}
        \caption{
        (a) Catalytic source sphere (S, gray) with radius $R_s=3.0$ and a colloidal sphere (C, red) with radius $R_c=4.0$, along with the $c_I$ concentration field (blue particles) near the source. The gray beads on the surface of the colloid signal that their enzymatic activities are inhibited by the $I$ particles. The repulsive potential parameters are $\epsilon_A$=2.0, $\epsilon_I$=0.1, and $\epsilon_B=0.2$. (b) The probability distribution function $P(r)$  of the distance of the colloid from the source. Three examples with different gate thresholds $c^T_I= 0.09, 0.3, 0.9$, are plotted. Inset: the evolution of distance $R_{SC}=\mid \textbf{r}_c-\textbf{r}_s \mid -(R_s+R_c)$ between the source and the colloid when $I$ is not an inhibitor.} \label{fig:INH-gate-sim}
\end{figure}

In the simulations the colloid and source spheres with radii $R_c$ and $R_s$, respectively, are contained in a slab-shaped volume with dimensions $L_x=L_y=50$ and $L_z=14$, with two confining parallel walls separated by a distance $L_z$. Periodic boundary conditions are used in the $x$ and $y$ directions. The initial separation between the S and C spheres is $R_{SC}^0$=17.7. The fixed sphere S acts as a source of species $I$ by catalyzing a reaction $A \to I$. The freely-moving colloid C is uniformly coated by an enzyme $E$ that catalyses the reaction $A \stackrel{E}{\rightarrow} B$. As in the previous section, there are fluid phase reactions, $I \stackrel{k^I_b}{\rightarrow} A $ and $B \stackrel{k^B_b}{\rightarrow} A $, that maintain the system in a nonequilibrium state. The S sphere produces inhomogeneous $A$ and $I$ concentration fields in its vicinity (see Fig.~\ref{fig:INH-gate-sim} (a)) given by
\begin{equation}
c_I(r) \propto \frac{1}{r}e^{-\kappa_I(r-R_s)},
\end{equation}
where $\kappa_I=\sqrt{k_b^I/D}$, with $D$ the solute diffusion coefficient, is the inverse screening length that determines how rapidly the $I$ concentration decays to zero in the bulk. A similar expression can be written for the $c_A(r)$ field. The energy parameters are chosen to satisfy $\epsilon_A > \epsilon_B > \epsilon_I$.

Since the catalyst $E$ is uniformly distributed over the entire surface of the colloid, in the absence of the source sphere and with a uniform supply of fuel $A$, the net propulsion force is zero by symmetry. When the $S$ sphere is present and the colloid enters its vicinity it will experience a depletion of $A$ fuel on its face that points towards the source. The subsequent reaction $A \stackrel{E}{\rightarrow} B$ converts A around the colloid to B. Since $\epsilon_B > \epsilon_I$ the diffusiophoretic force will cause the colloid to move towards the $S$ sphere and be captured by it, as shown in the inset in Fig.~\ref{fig:INH-gate-sim} (b).
\begin{figure}[htbp]
\centering
\resizebox{0.8\columnwidth}{!}{
      \includegraphics{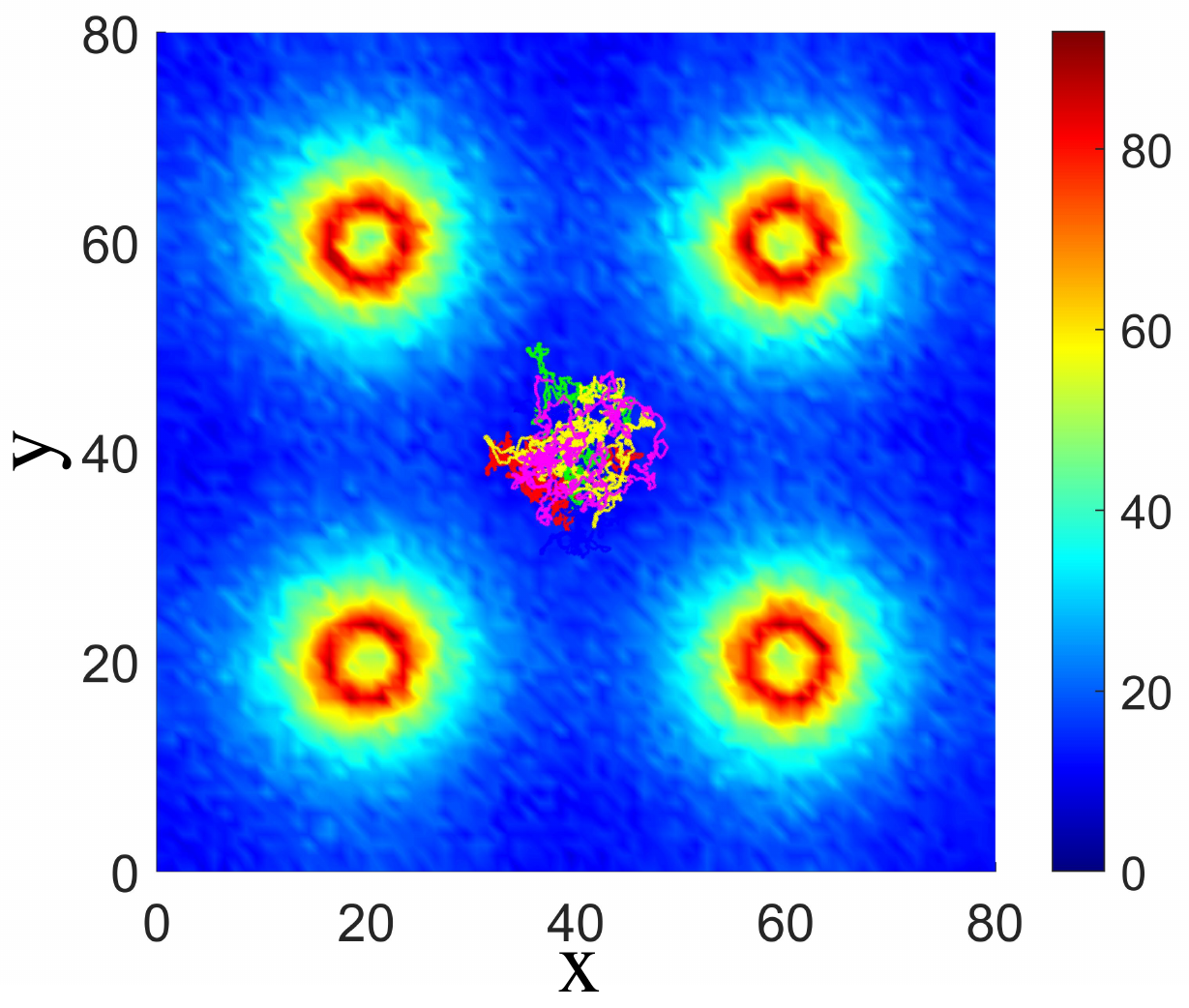}}
        \caption{
The concentration profile of species $I$ and trajectories of the colloid obtained from five independent realizations of the dynamics in a system with size $L_x=L_y=80$ and $L_z=14$ containing four source spheres with $c^T_I=0.3$.
} \label{fig:INH-gate-sim2}
\end{figure}

To avoid capture of colloid by the source, an INH gate is constructed on the colloid. Species $I$ acts  as the inhibitor. If the local concentration of $I$ around a enzyme bead, $c_{I}$, exceeds a pre-set threshold value, $c^T_{I}$, we suppose the reaction $A \stackrel{E}{\rightarrow} B$ is inhibited. If the value is either lower or higher than the threshold, the output of the gate is defined to be either 0 or 1, respectively. Therefore, when the colloid moves toward the source, the catalytic activities of enzymes facing the source are suppressed if $c_I>c^T_{I}$. In this circumstance the local concentration of $A$ is
greater than that on the face where the $A \to B$ reaction takes place and, since $\epsilon_A>\epsilon_B>\epsilon_I$, a diffusiophoretic force pushes the colloid away from the source. Figure~\ref{fig:INH-gate-sim} (b) shows the probability density of the colloid distance from the source, $P(r)=\langle\delta(\mid(\textbf{r}_C-\textbf{r}_S)\mid-r)\rangle$ where $\textbf{r}_C$ and $\textbf{r}_S$ are the center-of-mass positions of the colloid and source, respectively. The angle bracket denotes an average over time and realizations. When the INH gate is operative, one sees that the colloid cannot be captured by the source. Figure~\ref{fig:INH-gate-sim} (b) shows that for lower thresholds the colloid tends to remain at large distances from the source with quite sharp probability distributions, while for a large threshold it is able explore regions closer to the colloid with a broad distribution, without collapsing to a bound pair.

In Fig. 6 we consider a system with a colloid and four source spheres. The gate threshold is set to have an intermediate value, $c_I^T=0.3$. One sees that the colloid can move among the array of sources without touching them and, in fact, is trapped in a small region between the sources as a result of the action of the IHN gate.

\begin{figure}[htbp]
\centering
\resizebox{0.8\columnwidth}{!}{
      \includegraphics{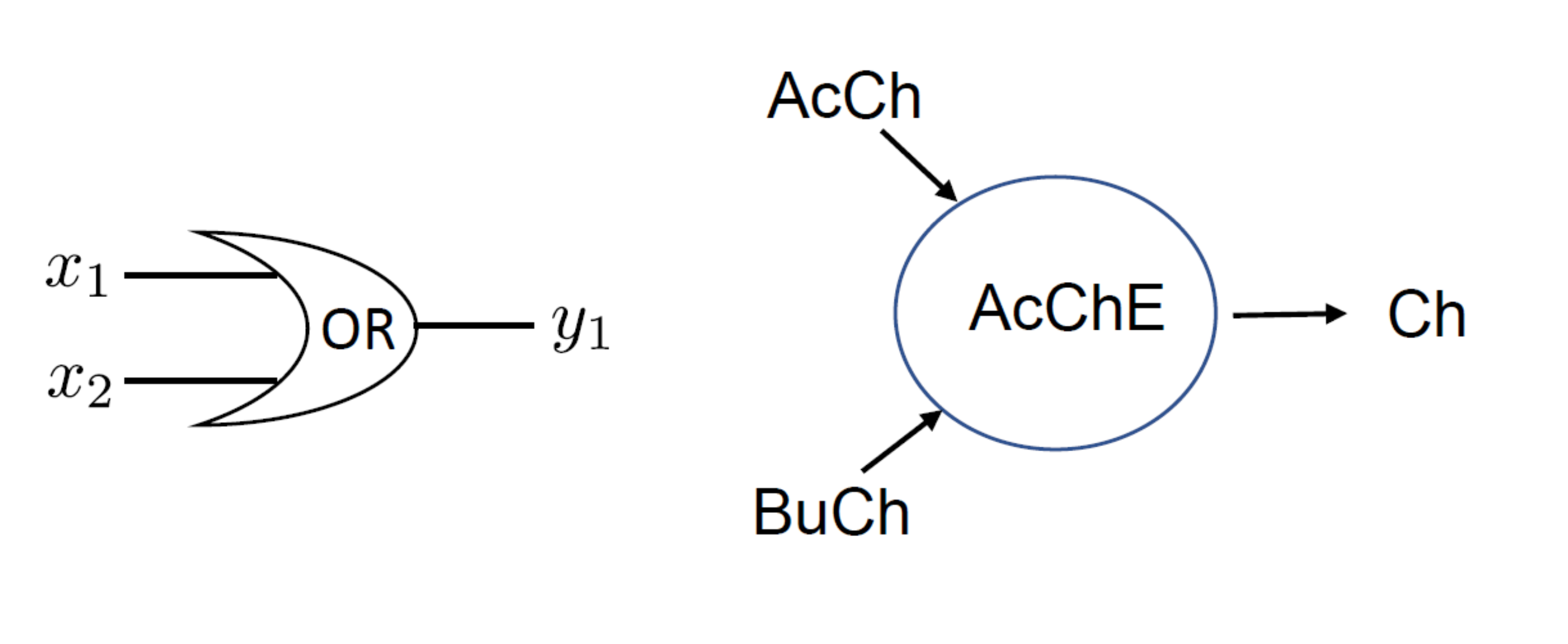}}
        \caption{
       An enzymatic reaction that simulates an OR gate. The enzyme acetylcholinesterase (AcCHE) catalyzes the decomposition of both acetylcholine (AcCH) and butyrycholine (BuCh) to the product choline (Ch). The inputs to the OR gate are $x_1={\rm AcCh}$ and $x_2={\rm BuCh}$ and the output is
       $y_1={\rm Ch}$.} \label{fig:OR-gate}
\end{figure}

\vspace{0.25cm} \noindent
{\sf OR gate:} An OR gate is shown in Fig.~\ref{fig:OR-gate} (a). This figure also shows an enzymatic implementation of such a gate that has been discussed in the literature.~\cite{KPS2017}  The single enzyme acetylcholinesterase (AcCHE) can catalyse the decomposition of both acetylcholine (AcCH) and butyrycholine (BuCh) to give the common product choline (Ch), so that it mimics the action of an OR gate.

The task using the OR gate is variation on that for the INH gate above but shows how a gate can be used to change the propulsion direction of a Janus colloid through a chemotactic effect in order to be captured by a source. We again have a source sphere that produces $I$ by consuming $A$. However, the colloid is Janus particle where one hemisphere $H_u$ is randomly covered by enzymes $E_1$ and $E_2$ with equal probability $1/2$, while $H_\ell$ is randomly covered by $E_1$ and inactive $E_0$ enzymes, also with probability $1/2$, as shown in fig. 8(a). The reaction on $E_1$ is  $A\underset{E_1}{\stackrel{\kappa_1}{\rightarrow}} B$ and that on $E_2$ is $I\underset{E_2}{\stackrel{\kappa_2}{\rightarrow}} B$. In our example the Janus colloid that supports two different enzymatic reactions functions acts an OR gate since the common product $B$ is produced if either $A$ or $I$ are inputs to the colloid.

Initially, the fluid has only A solute particles distributed uniformly. The potential parameters satisfy $\epsilon_A=\epsilon_I > \epsilon_B$. The Janus colloid uses the reaction on $E_1$ ($A\underset{E_1}{\stackrel{\kappa_1}{\rightarrow}} B$) for propulsion and, since $\epsilon_A > \epsilon_B$, the propulsion force is directed along $\hat{\bm{u}}$ pointing towards the $H_u$ cap. On time scales longer than the colloid orientational relaxation time, the active Janus colloid will undergo diffusive motion with an enhanced effective diffusion coefficient. If the reaction $A \to I$ does not take place so that the source sphere $S$ is inactive, with low probability the Janus colloid may encounter $S$ during its random walk. If the source converts $A \to I$ producing an inhomogeneous concentration field in its vicinity but the colloid reaction $I\underset{E_2}{\stackrel{\kappa_2}{\rightarrow}} B$ is not activated, i.e., only $A\underset{E_1}{\stackrel{\kappa_1}{\rightarrow}} B$ occurs on the colloid, the colloid still displays enhanced effective diffusion since $\epsilon_A=\epsilon_I$.

\begin{figure}[htbp]
\centering
\resizebox{1.05\columnwidth}{!}{
      \includegraphics{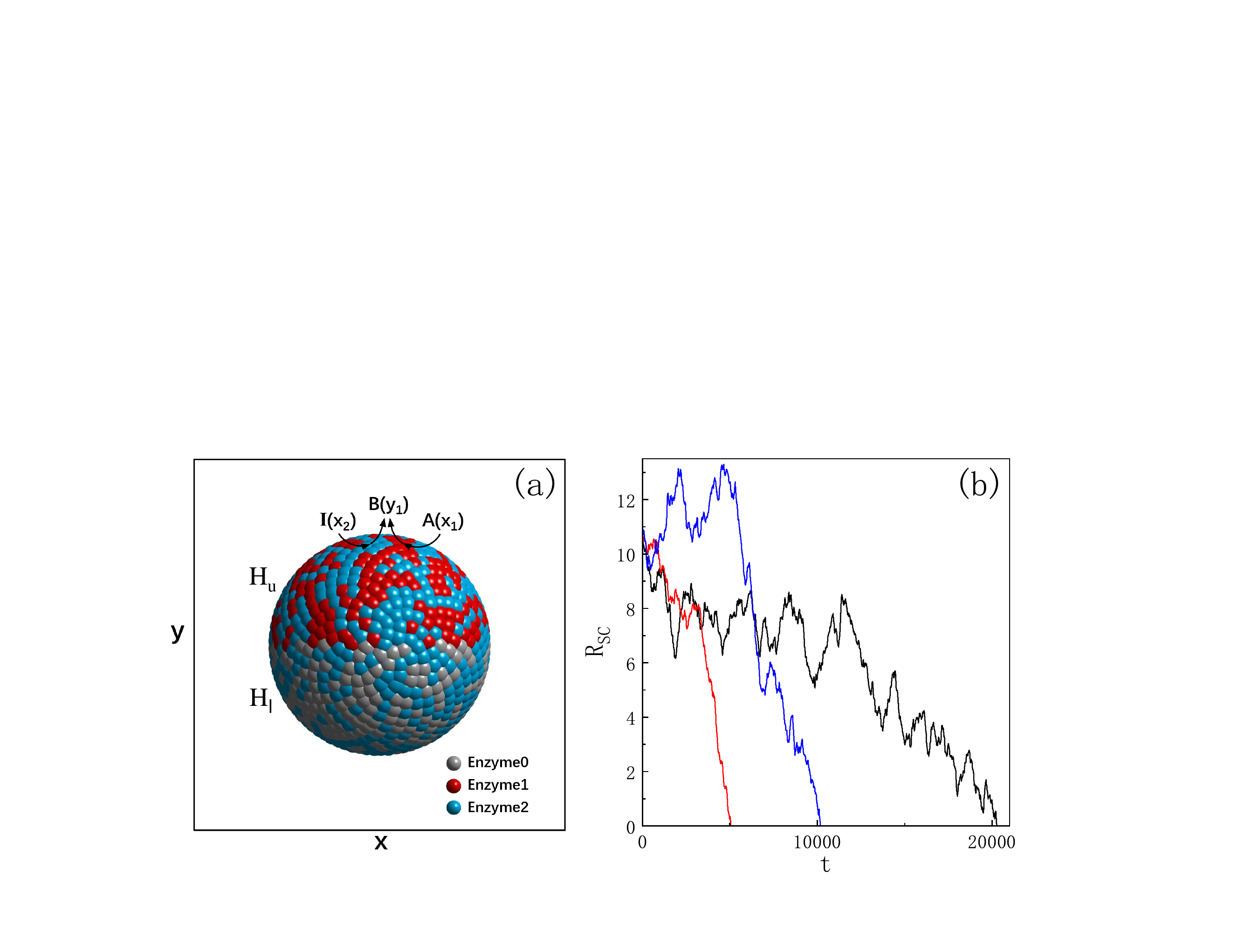}}
        \caption{
       (a) Janus colloid with equal numbers, $N^u_{E1}=N^u_{E2}=250$, of $E_1$ and $E_2$ enzymes randomly distributed on the upper $H_u$ hemisphere. Likewise, equal numbers, $N^{\ell}_{E1}=N^{\ell}_{E0}=250$, of $E_1$ and $E_0$ enzymes are randomly distributed on the lower hemisphere $H_\ell$. Both substrates $A$ and $I$ can be converted to $B$ by $E_1$ and $E_2$, respectively. Enzymes $E_0$ are chemically inactive. The interaction parameters are chosen to be $\epsilon_A=\epsilon_I=2.0$ and $\epsilon_B=0.1$. (b) Three examples of the evolution of the distance ($R_{SC}$) between the colloid and the source.
    } \label{fig:OR-simu}
\end{figure}

If the source is active and the colloid senses the $I$ concentration field, then the reaction  $I\underset{E_2}{\stackrel{\kappa_2}{\rightarrow}} B$ takes place on the $E_2$ sites that are uniformly distributed on the colloid (see Fig.~\ref{fig:OR-simu} (a)) simulating an OR gate. Now $B$ will also be preferentially produced on the face of the colloid that points towards the source, since species I has higher concentration near to the source (see Eq. (5)). In general this face will not point in the same direction as vector $\hat{\bm{u}}$. As a result the colloid diffusiophoretic force lies in a direction determined by the combined production of $B$ due to both catalytic reactions. This biases the propulsion direction towards the source, and leads to capture of the colloid. Figure~\ref{fig:OR-simu} (b) shows several stochastic trajectories that lead to capture by the source through this OR gate mechanism.

\section{Discussion}\label{sec:conc}
The results in this paper showed how small enzyme networks on colloidal motors could be used to construct a number of different chemical logic gates, and how these gates can enable the active colloids to perform simple tasks. The simulation results and continuum models provide detailed information on the way surface enzyme networks function to produce the inhomogeneous concentration fields that play an important part in how chemical gates function, how fuel is consumed in the network, and how the colloid is propelled.

The examples of sensing tasks presented above, while simple, show how an active colloid with chemical gates can use propulsion and sensing to change its dynamics to achieve a goal. Small enzyme networks of the sort shown in these examples can be used to construct all the logic gates, and strategies for constructing such gates have been described recently.~\cite{CKH20} The results lay the foundation for further research. The experiments on micromotors with a gated pH responsive DNA nanoswitch mentioned earlier are laboratory examples of how such gated active colloids could be used in applications.~\cite{PPJLSSRS19,Sanchez-gate2019}

There have been other studies that consider how active motion can be combined with logic functions or rules to yield complex dynamics; for example, model gates have been used in the context of active microfluidics~\cite{Woodhouse2017}, run-and-tumble particles with simple rules for gates~\cite{PLM2020}, and the use of chemical signals to communicate among self-propelled particles~\cite{Ziepke2022}. In the work presented here, model protein networks were used to build the chemical gates, and the simulations of gate operations were carried out at a coarse-grained microscopic level that takes into account the full reactive dynamics and fluid flows that underlie the propulsion mechanism and gate operations of the colloids, along with their interactions with their environments.

Proteins generally have a high specificity for substrate molecules enabling them to sense their temporally and spatially varying environments. Thus, the proteins on the surfaces of the colloids, either in chemical gates or acting singly, can be used to enable diverse sensing capability. Circuits built from these gates~\cite{KPS2017,PSSPK2008}, either on single colloids or collections of colloids, should be able to carry out computations needed to accomplish complicated tasks. The work presented here could be extended to treat these more complex and interesting situations, and may serve to interpret or suggest future experimental work on active colloids that sense and autonomously respond to their environments.

\appendix

\section{Continuum model}\label{app:cont-model1}

This Appendix gives some details of the calculations that enter Eqs.~(\ref{eq:colloid-prop-velocity}) and (\ref{eq:prop-velocity-bulk-fuel}). The enzymes with surface reactions in Eqs.~(\ref{eq:model1a}) and (\ref{eq:model1b}) have fractions $f_1^u=f_1^\ell=1/2$ and $f_2^u=1/2$ and $f_2^\ell=0$. Since the total global concentration is conserved, $c_A+c_B +c_I =c_0= {\rm const}$, it is convenient to consider the variables $c(\bm{r},t)$, $c_B(\bm{r},t)$ and $c_I(\bm{r},t)$ which satisfy the uncoupled equations
\begin{eqnarray}
\partial_t c(\bm{r},t) &=& D \nabla^2 c\\
\partial_t c_B(\bm{r},t) &=& D \nabla^2 c_B  - k^B_b c_B \\
\partial_t c_I(\bm{r},t) &=& D \nabla^2 c_I  - k^I_b c_I,
\end{eqnarray}
where $k_D=4 \pi D R$. These fields are subject to the following boundary conditions on the surface of the colloid:
\begin{eqnarray}
k_D R \partial_r c(r,\theta,t)\mid_R &=& 0, \\
k_D R \partial_r c_B(r,\theta,t)\mid_R &=& -k_2 \Theta_u c_I(R,\theta,t),\\
k_D R \partial_r c_I(r,\theta,t)\mid_R &=& (k_2 \Theta_u +k_1)  c_I(R,\theta,t) \nonumber \\
&&+k_1 c_B(R,\theta,t) -k_1 c_0,
\end{eqnarray}
where $\Theta_u$ is a Heaviside function that is unity on the upper hemisphere and zero otherwise, while at infinity we have
$\lim_{r \to \infty} c(r,\theta,t) =c_0$, $\lim_{r \to \infty} c_B(r,\theta,t) =0$ and $\lim_{r \to \infty} c_I(r,\theta,t) =0$.

Henceforth we consider the steady state versions of these equations. The steady state solutions of the fluid phase reaction-diffusion can be written as
\begin{eqnarray}
c(r,\theta) &=& c_0 + \sum_{\ell =0}^\infty a_\ell^c \Big(\frac{R}{r}\Big)^{\ell+1} P_\ell(\cos \theta) \\
c_B(r,\theta) &=& \sum_{\ell =0}^\infty a_\ell^B f_\ell^B(r) P_\ell(\cos \theta)\\
c_I(r,\theta) &=& \sum_{\ell =0}^\infty a_\ell^I f_\ell^I(r) P_\ell(\cos \theta),
\end{eqnarray}
where, for $\alpha=B,I$,
\begin{equation}
f_\ell^\alpha(r)=\frac{\sqrt{\nu_b^\alpha R}\; K_{\ell+\frac{1}{2}}(\nu_b^\alpha r)}{\sqrt{\nu_b^\alpha r}\; K_{\ell+\frac{1}{2}}(\nu_b^\alpha R)},
\end{equation}
with $\nu_b^\alpha = \sqrt{k_b^\alpha/D}$ and $K_n(x)$ an associated Bessel function of the second kind. The boundary condition for $c$ is
\begin{equation}
\partial_r c(r,\theta)\mid_R=-k_D R \sum_{\ell=0}^\infty a_\ell^c \frac{(\ell+1)}{R} P_\ell(\cos \theta)=0,
\end{equation}
from which we conclude that $a_\ell^c=0$ for all $\ell$. Therefore $c(r,\theta)=c_0$ as expected. The boundary condition for $c_B$ yields
\begin{equation}\label{eq:cB-BC}
\frac{2 Q_\ell^B}{2\ell +1} a_\ell^B =  \frac{k_2}{k_D} \sum_{m=0}^\infty N_{\ell m} a_m^I ,
\end{equation}
where $x=\cos \theta$, $Q_\ell^\alpha$ for $\alpha = B,I$ is defined by,
\begin{equation}\label{eq:Q}
Q_\ell^\alpha = \frac{(\nu_b^\alpha R) \; K_{\ell+\frac{3}{2}}(\nu_b^\alpha R)}{ K_{\ell+\frac{1}{2}}(\nu_b^\alpha R)} -\ell ,
\end{equation}
and the matrix element $N_{\ell m}=\int_0^1 dx \; P_\ell(x) P_m(x)$ can be computed analytically. After some rearrangement of the $c_B$ and $c_I$ boundary conditions we can obtain a closed linear equation for the $a_\ell^I$ coefficients:
\begin{equation}
a_\ell^I = 2\frac{k_1}{k_D}c_0 \Big(\bm{\mathcal{M}}^{-1} \Big)_{\ell 0},
\end{equation}
where the elements of $\bm{\mathcal{M}}$ are
\begin{equation}\label{eq:M-model-1}
\mathcal{M}_{\ell m} =\frac{2 }{2\ell +1}\Big(Q_\ell^I+\frac{k_1}{k_D}\Big) \delta_{\ell m}+
\frac{k_2}{k_D} \Big(1+\frac{k_1}{k_D}\frac{1}{Q_\ell^B}\Big) N_{\ell m}.
\end{equation}
Once these $a_\ell^I$ coefficients are obtained, the coefficients $a_\ell^B$ can be found after substitution into Eq.~(\ref{eq:cB-BC}); thus, the concentration fields $c_I(r,\theta)$, $c_I(r,\theta)$ and $c_A(r,\theta)$ can be obtained.

Next, these results will be used to compute the propulsion velocity of the colloid. The expression for the diffusiophoretic velocity is given by~\cite{GK18,GK18a}
\begin{equation}\label{eq:colloid-velocity}
\bm{V}_u =V_u \hat{\bm{u}} =\frac{1}{1+2b/R} \sum_\alpha b_\alpha \overline{\bm{\nabla}_s c_\alpha(\bm{r})}^S,
\end{equation}
where $\alpha=A,I,B$ and, as noted earlier, the overline denotes an average over the surface of the colloid, $\overline{O}^S=\frac{1}{4\pi R^2}\int d\bm{r} \; \delta(r-R) O $,
and $b_\alpha =\frac{k_B T}{\eta} (K_\alpha^{(1)}+b K_\alpha^{(0)})$
with $b$ the slip length and
\begin{equation}
K_\alpha^{(n)}=\int_R^{R+r_c} dr \; (r-R)^n \Big(e^{-\beta V_{\alpha c}} -1 \Big),
\end{equation}
and $r_c$ the distance beyond which the solute-colloid interactions vanish.

For our system we have (using $c_A=c_0 -c_I -c_B$)
\begin{eqnarray}
V_u&=& \frac{1}{1+2b/R} \frac{k_B T}{\eta}\Big[\Lambda_{IA} \overline{ \hat{\bm{u}} \cdot \bm{\nabla}_s c_I(R,\theta)}^S \nonumber\\
&& \quad+\Lambda_{BA} \overline{ \hat{\bm{u}} \cdot \bm{\nabla}_s  c_B(R,\theta)}^S  \Big],
\end{eqnarray}
where $\Lambda_{\alpha \alpha'} = \Lambda^{(1)}_{\alpha \alpha'} +b\Lambda^{(0)}_{\alpha \alpha'}$ with
\begin{equation}
\Lambda^{(n)}_{\alpha \alpha'}=\int_R^{R+r_c} dr \; (r-R)^n \Big(e^{-\beta V_{\alpha c}} -e^{-\beta V_{\alpha' c}} \Big).
\end{equation}
This is Eq.~(\ref{eq:colloid-prop-velocity}) in the text with $b=0$ for velocity stick boundary conditions. Since $\hat{\bm{u}} \cdot \bm{\nabla}_s = - \frac{\sin \theta}{r} \partial_\theta$. This equation can also be written as
\begin{eqnarray}\label{eq:Vd-simple-model-2}
V_u&=&-\frac{1}{1+2b/R}\frac{k_B T}{R\eta} \Big[\Lambda_{IA} \;\overline{ \sin \theta \; \partial_\theta c_I(R,\theta)}^S \nonumber \\
&&+ \Lambda_{BA}\;\overline{ \sin \theta \; \partial_\theta c_B(R,\theta)}^S  \Big],
\end{eqnarray}
Integrating by parts in the surface averages and substituting the expression for $c_I(R,\theta)$ and $c_B(R,\theta)$ we get the final result for the colloid velocity,
\begin{eqnarray}\label{eq:Vd-simple-model}
V_u&=& -\frac{2}{1+2b/R}\frac{k_B T}{R\eta} \frac{k_1}{k_D}c_0  \\
&& \times \sum_{m=0}^\infty \Big[\frac{2}{3} \Lambda_{IA} \;\delta_{m1}
+ \frac{k_2}{k_D}\frac{1}{Q^B_1} \Lambda_{BA}\;N_{1m}  \Big] \Big(\bm{\mathcal{M}}^{-1} \Big)_{m 0}.\nonumber
\end{eqnarray}

\vspace{0.25cm}
\noindent
{\sf Direct supply of species $I$ from fluid phase}: We can compare the above results with a system where fuel $I$ is supplied in the bulk:
\begin{eqnarray}
I \stackrel{\kappa_2}{\rightarrow} B , &\quad& {\rm on \; hemispherical \; surface}\label{eq:bulkIa}\\
B \stackrel{k^B_b}{\rightarrow} I , &\quad& {\rm in \; the \; fluid \; phase},\label{eq:bulkIb}
\end{eqnarray}
The reaction-diffusion equations are
\begin{eqnarray}
\partial_t c_B(\bm{r},t) &=& D \nabla^2 c_B  - k^B_b c_B \\
\partial_t c_I(\bm{r},t) &=& D \nabla^2 c_I  + k^B_b c_B,
\end{eqnarray}with boundary conditions
\begin{eqnarray}
k_D R \partial_r c_B(r,\theta,t)\mid_R &=& -k_2 \Theta_u c_I(R,\theta,t),\\
k_D R \partial_r c_I(r,\theta,t)\mid_R &=& k_2 \Theta_u c_I(R,\theta,t),
\end{eqnarray}
while at infinity we have
$\lim_{r \to \infty} c_B(r,\theta,t) =0$ and $\lim_{r \to \infty} c_I(r,\theta,t) =c_0$.

Similar to the calculation given above, the solutions to the reaction-diffusion equations can be written as
\begin{equation}
c_B(r,\theta) = \sum_{\ell =0}^\infty \tilde{a}_\ell^B f_\ell^B(r) P_\ell(\cos \theta),
\end{equation}
with $c_I(r,\theta)=c_0-c_B(r,\theta)$ that follows from number conservation. Determining the coefficients from the boundary conditions gives
\begin{equation}
\tilde{a}_\ell^B =\sum_{m=0}^\infty \Big( \tilde{\bm{\mathcal{M}}}^{-1}  \Big)_{\ell m} \tilde{\mathcal{E}}_m,
\end{equation}
where $\tilde{\mathcal{E}}_m=(k_2/k_D)c_0\int_0^1 dx P_m(x)$ and
\begin{equation}
\tilde{\mathcal{M}}_{\ell m}=\frac{2 Q_\ell^B}{2\ell +1} \delta_{\ell m}+
\frac{k_2}{k_D}  N_{\ell m}.
\end{equation}
Using this result the motor velocity is
\begin{equation}\label{eq:model2}
V_u= \frac{1}{1+2b/R}\frac{k_B T}{R\eta} \frac{2}{3} \Lambda_{IB}  \tilde{a}_1^B.
\end{equation}

\section{Simulation model and parameters}\label{app:sim-model1}
The active colloid with radius $R=4.0$ and mass $M_c$ is constructed from spherical beads linked by stiff harmonic bonds to insure that the spherical shape of the colloid is maintained during the evolution of the system. The $N_E$=1000 surface beads with radius $\sigma=1.0$ are coarse-grained enzymatic sites. The colloid is contained in a cubic simulation box with size $V=L_x\times L_y\times L_z$ containing reactive particles. Periodic boundary conditions are applied.

The point-like substrate molecules with mass $m$ interact with the surface enzyme beads through repulsive Lennard-Jones (LJ) potentials, $V_{LJ}(r)=4\epsilon_{\alpha}((\sigma/r)^{12}-(\sigma/r)^{6}+1/4)\Theta(r_c-r)$,
where $\Theta$ is a Heaviside function, $\alpha$ labels species, and $r_c = 3+2^{1/6}=4.125$ is the cutoff distance beyond which the potential is zero.

For simulation where the colloid is confined to the center plane in a simulation box with a slab geometry, five beads are selected: one at the center-of-mass of the colloid, and others on a circle with $r=3.5$ in the $x$-$y$ plane at $z=L_z/2$. They interact with the walls at $z=0$ and $z=L_z$ through a 9-3 LJ potential, $V_{LJ}^{93}(r)=\epsilon_w [(\sigma_w/r)^{9}-(\sigma_w/r)^3 ]$, where $\epsilon_w$ and $\sigma_w$ are the wall energy and distance parameters, respectively. The interaction between the fluid particles and the source sphere are 6-12 repulsive LJ potentials with $\epsilon=0.1$.

Reactive events that convert substrate to product take place on the enzyme beads $E_\nu$ ($\nu$ labels different enzymes) with probabilities $p^R_\nu$.~\cite{Huang_etal_2018}
For the continuum calculation, the reaction rate constant for the colloid can be estimated from the kinetic theory expression,
\begin{equation}\label{eq:rate-coef}
k_\nu=p^R_\nu r^2_c \sqrt{8\pi k_BT/\mu_{cs}},
\end{equation}
and the reduced mass is $\mu_{cs}=(M_c m)/(M_c+m)$.

The system is evolved using hybrid molecular dynamics-multiparticle collision (MPC) dynamics.~\cite{MK99,MK00,Ray08,GIKW09} Simulation results are reported in dimensionless units based on energy $\epsilon$, mass $m$, and cell length $a_0$. The time is in units of $t(m a_0^2/\epsilon)^{1/2}\rightarrow t$, distance parameter $r/a_{0}\rightarrow r$, and temperature $k_BT/\epsilon \rightarrow T$. The system temperature $T$ is $\frac{1}{6}$. The average number of fluid particles per cell is $c_0=10.2$. The mass of the colloid is given by $M_c=\frac{4}{3}\pi R^3c_0$, so that the colloid is approximately neutrally buoyant. The MPC rotation angle is $\phi=\frac{\pi}{2}$ and collision time interval is $\tau_{MPC}=0.5$. The velocity Verlet algorithm is used to integrate the Newton's equation of motion with $\tau_{MD}=0.005$. The Schmidt number is $S_c=1.39$. Other parameters: harmonic spring force constant $k_s$=60, wall $\epsilon_w=5.0$, $\sigma_w=L_z/2$,  fluid rate constants $k_b^B=k_b^I=0.001$, common diffusion coefficient $D=0.097$, viscosity from MPC expression $\eta=1.35$. The values of $\epsilon_{\alpha}$ are given in the text. Grid shifting is employed to ensure Galilean invariance.~\cite{Ihle_Kroll_01}

\bibliography{references-smart}

\begin{thebibliography}{68}
\bibitem [{sti()}]{stick}%
  \BibitemOpen
  \href@noop {} {}\bibinfo {note} {This formula applies for stick boundary
  conditions for the fluid velocity field on the colloid surface. The more
  general expression for arbitrary slip is given in the Appendix.}\BibitemShut
  {Stop}%
\end{thebibliography}

\textbf{Acknowledgements}
We thank Liyan Qiao for useful discussions. This work is supported by the National Natural Science Foundation of China (No.: 12274110, 11974094) and the Natural Sciences and Engineering Research Council of Canada.

\textbf{Author contributions}
J. C. and R. K. contributed at all stages of this work. J. Q. performed part of simulation.

\textbf{Competing interests}: The authors declare no competing financial interests.

\end{document}